\documentclass{report} 
\usepackage{graphicx}

\def\msunyr             {M$_{\odot}$~yr$^{-1}$}
\def\kms       {~km~sec$^{-1}$}
\def\deg{^{\circ}}

\renewcommand{\abstract}[1]{{ \footnotesize \noindent {\bf Abstract} #1\\}}

\renewcommand{\author}[1]{\subsection*{#1}}
\newcommand{\address}[1]{\subsection*{\it#1}}

\begin{document}

\chapter*{Galaxy Clusters with Chandra\footnote{Contribution to XIII
Rencontres de Blois 2001, ed. L. M. Celnikier}}
\author{W. Forman$^1$, C. Jones$^1$, M. Markevitch$^{1,3}$, A. Vikhlinin$^{1,3}$, \&
E. Churazov$^{2,3}$}


\address{1) Smithsonian Astrophysical Observatory, Cambridge, MA, USA\\
2) MPI fur Astrophysik, Garching, Germany\\
3) Space Research Institute, Moscow, Russia}

\abstract{We discuss Chandra results related to 1) cluster mergers and
cold fronts and 2) interactions between relativistic plasma and hot
cluster atmospheres. We describe the properties of cold fronts using
NGC1404 in the Fornax cluster and A3667 as examples. We discuss
multiple surface brightness discontinuities in the cooling flow cluster
ZW3146. We review the supersonic merger underway in CL0657. Finally,
we summarize the interaction between plasma bubbles produced by AGN
and hot gas using M87 and NGC507 as examples.}

\section{Cluster Formation at High Angular Resolution}

For many years clusters were thought to be dynamically relaxed systems
evolving slowly after an initial, short-lived episode of violent
relaxation. However, while the dynamical timescale for the richest
cluster cores is comfortably less than the Hubble time, other less
dense clusters have dynamical timescales comparable to or longer than
the age of the Universe~\cite{wrfgunn1972}. X-ray images, starting with
Einstein and continuing with ROSAT and ASCA, and now with Chandra and
XMM-Newton, provide a powerful technique to ``map'' the structure in
the gravitational potential of massive systems containing hot gaseous
atmospheres. 

The X-ray observations supported the now prevalent idea that structure
in the Universe has grown through gravitational amplification of small
scale instabilities or hierarchical clustering. At one extreme, some
clusters grow, in their final phase, through mergers of nearly equal
mass components. Such mergers can be spectacular events involving
kinetic energies as large as $\sim 10^{64}$ ergs, the most energetic
events since the Big Bang.  More common are smaller mergers and
accretion of material from large scale filaments.  The ROSAT image of
A85 shows the relationship between large scale structure and cluster
merging where small groups are detected infalling along a filament
into the main cluster~\cite{wrfdurret1998}. Chandra and XMM with high
angular resolution and high throughput provide new insights into
cluster formation and evolution.

\paragraph{Cluster Cold Fronts}

Prior to the launch of Chandra, sharp gas density discontinuities had
been observed in the ROSAT images of A2142 and
A3667~\cite{wrfmark1999}. Since both clusters exhibited characteristics
of major mergers, these features were expected to be shock
fronts. However, the first Chandra observations showed that these were
not shocks, but a new kind of structure -- cold
fronts~\cite{wrfmark2000}. Their study has provided new and detailed
insights into the physics of the intracluster medium (ICM)~\cite{wrfvik2001a,wrfvik2001b}.

\begin{figure}
\centerline{\includegraphics[width=0.5\textwidth,bb=90 185 500 610,clip]{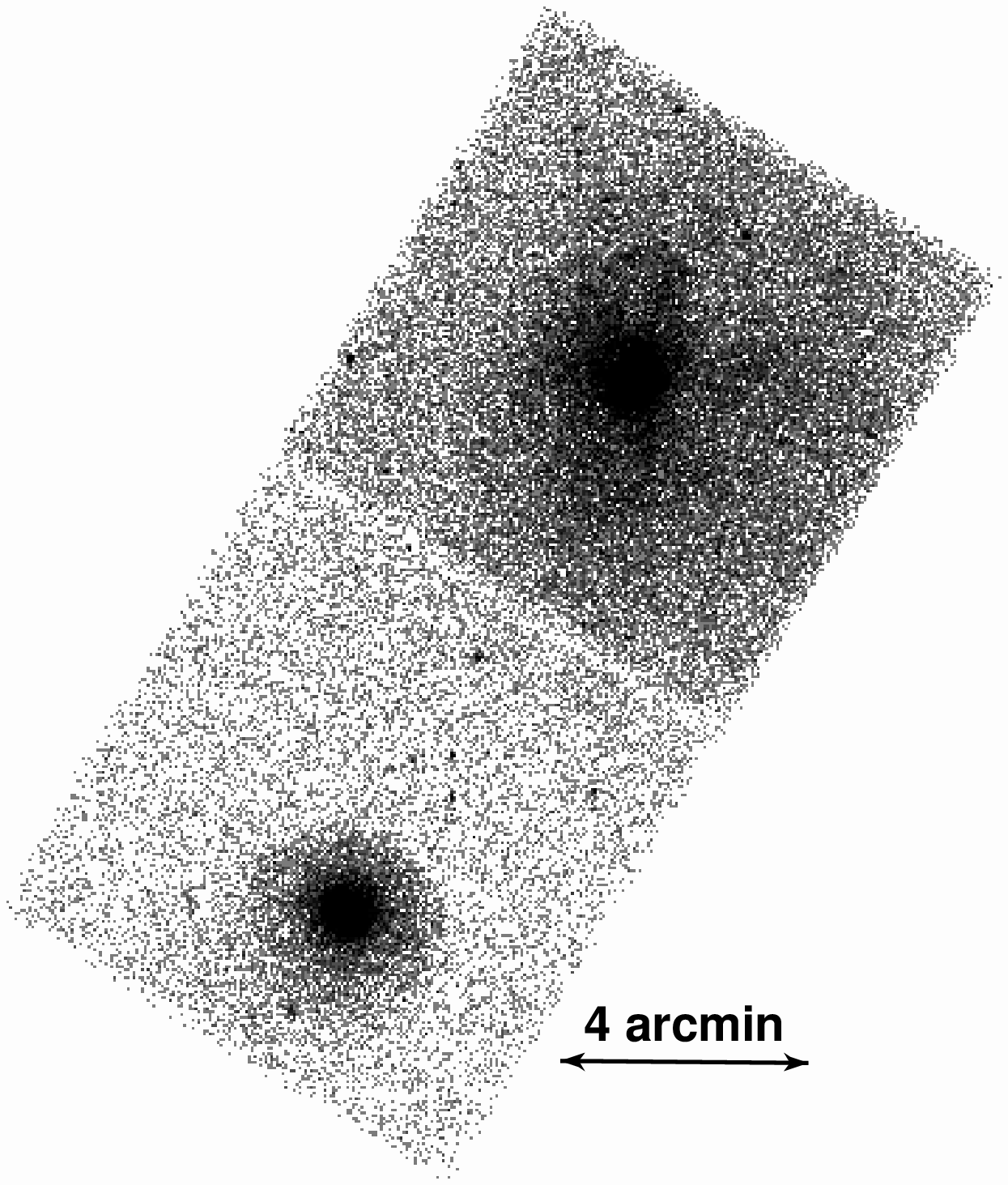}
\includegraphics[width=0.5\textwidth,bb=111 200 464 549,clip]{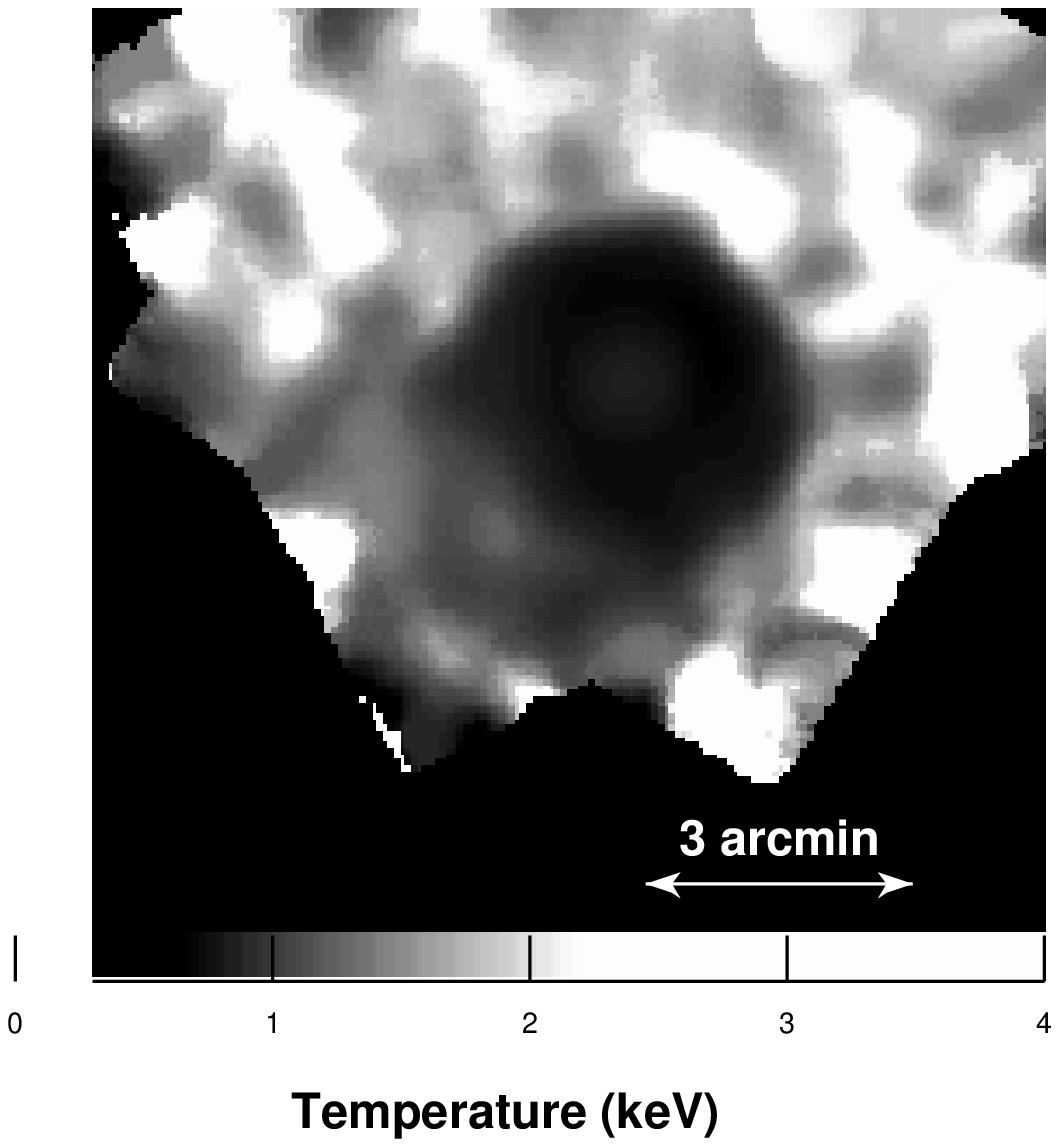}}
\caption{The ACIS observation of NGC1404 and NGC1399. (\textbf{a}) 
  shows the 0.5--2.0 keV band image of the Fornax cluster.  The gas
  filled dark halo surrounding NGC1404 is at the lower left
  (southeast) while the cluster core, dominated by the halo surrounding
  NGC1399 lies at the upper right (northwest). (\textbf{b}) The
  temperature map of the Fornax region. The cold core surrounding
  NGC1404 has a temperature of less $\sim1$ keV while the
  surrounding gas has a temperature of 1.5 keV. }
\label{wrfn1404_image}
\label{wrfn1404_tmap}
\end{figure}

Fig.~\ref{wrfn1404_image}a shows a striking example of a cold front as
NGC1404 and its gaseous corona approaches the cluster center (to the
northwest)~\cite{wrfanil2002}. The image clearly shows the sharp edge of the surface
brightness discontinuity, shaped by the ram pressure of the cluster
gas.  The temperature map (Fig.~\ref{wrfn1404_tmap}b) confirms that the
infalling cloud is cold compared to the hotter Fornax
ICM.

\paragraph{Physics of Cold Fronts}

The best studied cold front is that in A3667 ($z=0.055$).  Although
expected to exhibit a shock front based on its ROSAT
image~\cite{wrfmark1999}, the sharp feature observed in the Chandra
observation is the boundary of a dense
cold cloud, a merger remnant~\cite{wrfvik2001a,wrfvik2001b}. The surface
brightness profile, converted to gas density, and the gas temperature
distribution yield the gas pressure on both sides of the cold front.
The difference between the two pressures is a measure of the ram
pressure of the ICM on the moving cold front.  Hence, the precise
measurement of the gas parameters yields the cloud velocity.  The
factor of two difference in pressures between the free streaming
region and the region immediately inside the cold front yields a Mach
number of the cloud of $1\pm0.2$ ($1430\pm290$ km
s$^{-1}$)~\cite{wrfvik2001a}.

In addition to the edge, a weak shock is detected.  The distance
between the cold front and the weak shock ($\sim350$ kpc) and the
observed gas density jump at the shock (a factor of 1.1-1.2) yield the
shock's propagation velocity, $\sim1600$ km s$^{-1}$, which is consistent with
that derived independently from the pressure jump across the cold
front~\cite{wrfvik2001a}.

The A3667 observation provides important information on the efficiency
of transport processes in clusters. As the surface brightness profile
shows (see \cite{wrfvik2001a}), the density ``edge'' is very
sharp. Quantitatively, Vikhlinin et al. found that the width of the
front was less than $3.5''$ (5 kpc). This sharp edge requires that
transport processes across the edge be suppressed, presumably by
magnetic fields. Without such suppression, the edge should be broader
since the  electron Coulomb mean free path is about 13
kpc, several times the width of the cold front~\cite{wrfvik2001a}.
Furthermore, Vikhlinin et al. observed that the cold front appears
sharp only over a sector of about $\pm30\deg$ centered on the
direction of motion, while at larger angles, the sharp boundary
disappears~\cite{wrfvik2001b}. The disappearance can be explained by the
onset of Kelvin-Helmholtz instabilities, as the ambient ICM gas flows
past the moving cold front. To explain the observed extent of the
sharp boundary, the instability must be partially suppressed, e.g., by
a magnetic field parallel to the boundary with a strength of
$7-16\mu$G with a corresponding pressure of only 10-20\% of the
thermal pressure~\cite{wrfvik2001b}.

\begin{figure}
\centerline{\includegraphics[width=0.450\textwidth,bb= 111 270 477
520,clip]{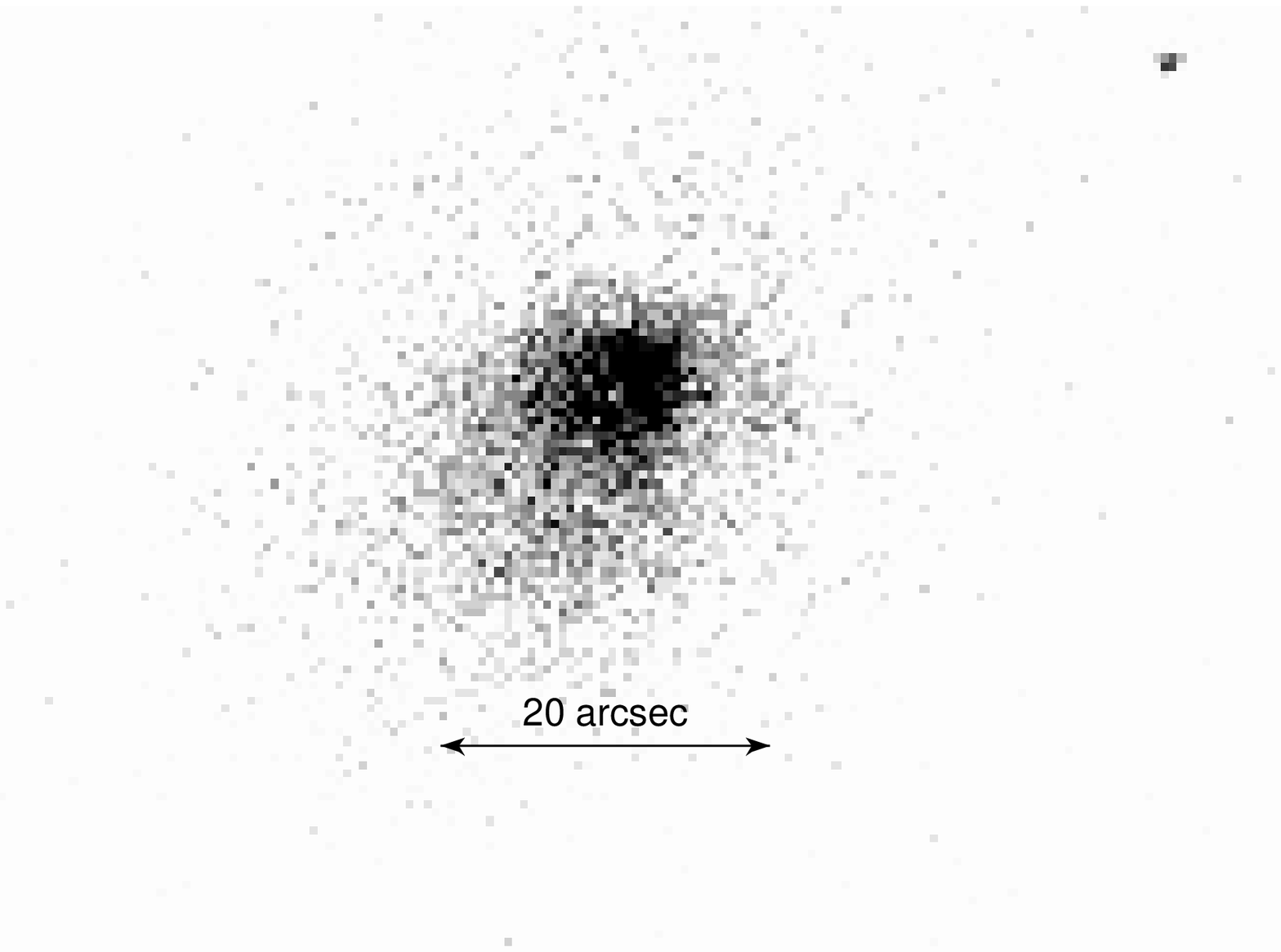}\includegraphics[width=0.450\textwidth,bb= 63
220 500 500,clip]{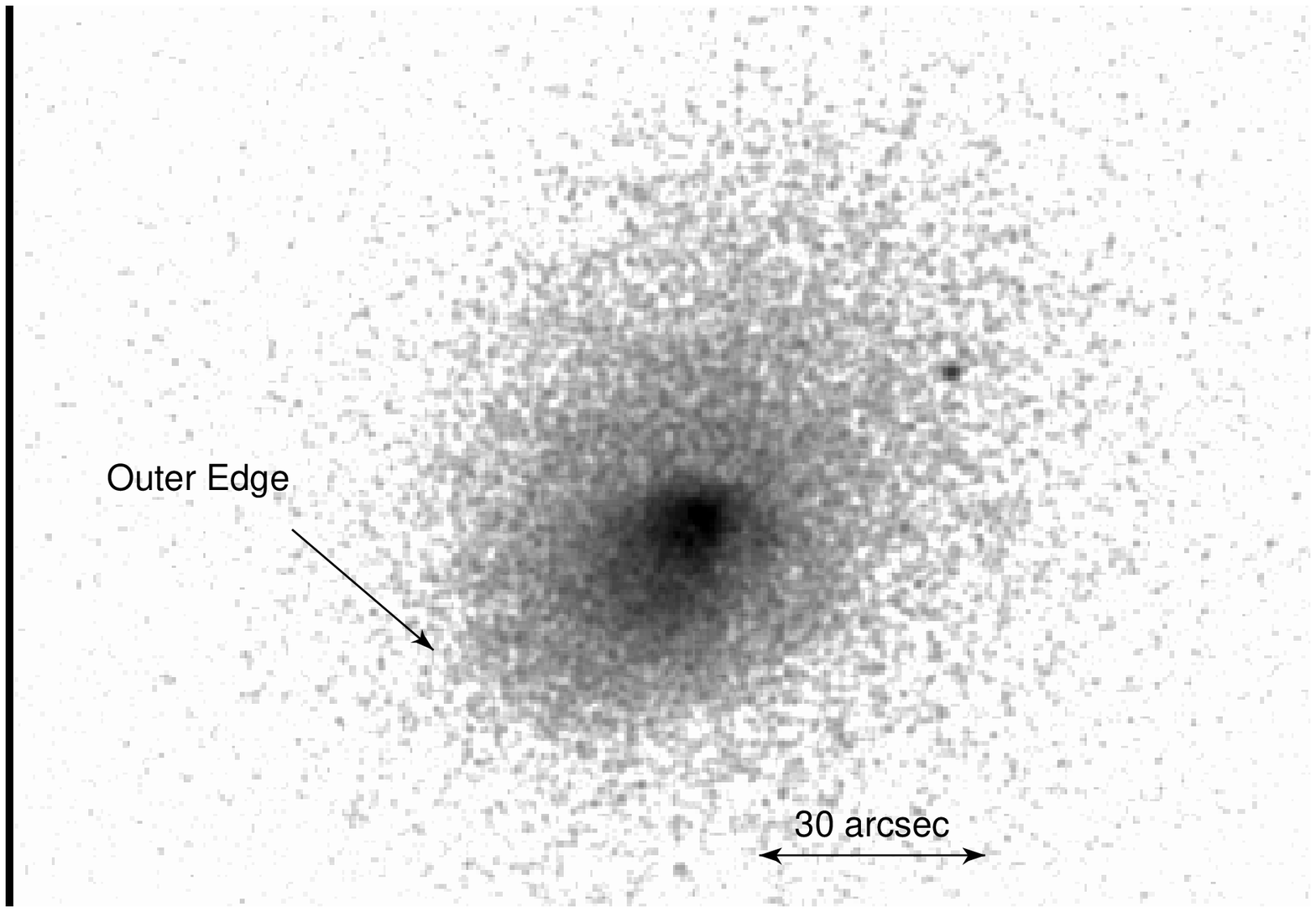}}
\caption{Multiple X-ray surface brightness edges in ZW3146 ($z=0.296$). 
  (\textbf{a}) The 0.5-2.0 keV image of the central
  region of ZW3146 shows the two inner edges at $3''$ and $8''$.
  (\textbf{b}) The right panel shows the edge at $35''$. }
\label{wrfzw3146}
\end{figure}

\paragraph{ZW3146 -- Multiple ``Fronts''}

ZW3146 is a moderately distant ($z=0.2906$; 5.74 kpc per arcsec)
cluster with a remarkably high mass deposition rate that was estimated
to exceed 1000 \msunyr~\cite{wrfedge}. The Chandra image further
demonstrates the remarkable nature of this cluster -- on scales from
$3''$ to $30''$ ($\sim20$ kpc to 170 kpc), three separate X-ray
surface brightness edges are detected (see Fig.~\ref{wrfzw3146} and
Forman et al.~\cite{wrfmoriond}). At the smallest radii, two edges are
seen to the northwest and north of the center (see Fig.\ref{wrfzw3146}a).
The first, at a radius of $\sim3''$ (17 kpc), spans an angle of nearly
$180\deg$ with a surface brightness drop of almost a factor of 2.  The
second edge, at a radius of $\sim8''$ (45 kpc) spans only $90\deg$ but
has a surface brightness drop of almost a factor of 4.  The third edge
(see Fig.~\ref{wrfzw3146}b) lies to the southeast, about $35''$ (200 kpc)
from the cluster center, has a decrease of about a factor of 2, and,
as with the first edge, extends over an angle of almost $180\deg$.

The variety of morphologies and scales exhibited by these sharp edges
or cold fronts is quite remarkable. Possibly the edges may arise from
moving cold gas clouds that are the remnants of merger activity as
observed in A2142 and A3667 or as oscillations (or ``sloshing'') of
the cool gas at the center of the cluster potential as observed in
A1795~\cite{wrfa1795}. The extremely regular morphology of ZW3146 on
large linear scales seems to exclude a recent merger and, hence,
``sloshing'' of the gas seems the more likely explanation for the
observed edges.  High resolution, large scale structure simulations
show that dense halos, formed at very early epochs, survive
cluster collapse~\cite{wrfghigna1,wrfghigna2}. While most of
the dark matter halos, having galaxy size masses, are associated with
the sites of galaxy formation, the larger mass halos also may survive
(without their gas) or may have fallen into the cluster only
recently. Hence, we might expect to find a range of halo mass
distributions moving within the cluster potential. We speculate that,
as these halos move, the varying gravitational potential could accelerate
the cool dense gas that has accumulated in the cluster core and could
produce the ``sloshing'' needed to give rise to the multiple surface
brightness edges observed in some clusters.

\begin{figure} [!ht]
\centerline{\includegraphics[width=0.50\textwidth]{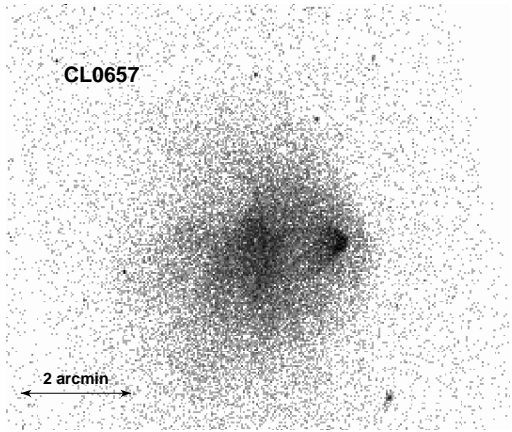}}
\caption{The Chandra image of the cluster CL0657.  The cluster
  exhibits the classic properties of a supersonic merger -- a dense
  (cold) ``bullet'' traversing the hot cluster with a leading shock front (Mach
  cone). The gas parameters across the front imply the cold core is
  traversing the cluster at a supersonic velocity with a Mach number
  of $M\sim2-3$. The disrupted core of the cluster can be seen to the
  east of the ``bullet''.
}
\label{wrf0657}
\end{figure}

\paragraph{CL0657 -- A Prototypical Cluster Shock Front}

CL0657 ($z=0.296$) was discovered by Tucker et al. as part of a search
for ``failed'' clusters, clusters that were X-ray bright but had
few, if any, optical galaxies~\cite{wrftucker1}. From ASCA observations,
this cluster was found to have a remarkably hot gas temperature of
about 17 keV, making it the hottest cluster known~\cite{wrftucker2}.

The Chandra image of CL0657 shows the classic properties of a
supersonic merger (see Markevitch et al. for a detailed discussion of
this cluster~\cite{wrfmaxim0657}).  We see a dense (cold) core moving
to the west after having traversed, and disrupted, the core of the
main cluster. Leading the cold, dense core is a density discontinuity
which appears as a shock front (Mach cone). The spectral data show
that the gas to the east (trailing the shock) has been heated by the
passage of the shock. The detailed gas density parameters confirm that
the ``bullet'' is moving to the west with a velocity of 3000--4000
\kms, approximately 2-3 times the sound speed of the ambient
gas. CL0657 is the first clear example of a relatively strong shock
arising from cluster mergers.

\section{The Radio---X-ray Connection -- or Bubbles, Bubbles Everywhere}

Prior to the launch of Chandra, ROSAT observations of NGC1275 and M87
provided hints of complex interactions between radio emitting plasmas
ejected from AGN within the nuclei of dominant, central cluster
galaxies~\cite{wrfboh1993,wrfboh1995,wrfchur2000,wrfchur2001}. With the launch of Chandra,
the interaction between relativistic plasma, with and without detected
radio emission, and the hot intracluster medium (ICM) has been
observed in many systems and is now a major area of
investigation~\cite{wrffabian2000,wrfmcnamara2000,wrfbubbles,
wrffinoguenov2001,wrfvrtilek2001,wrfmazzotta2001,wrfjones,wrfnulsen,wrfmcnamara}.

\paragraph{M87 and the Evolution of Buoyant Plasma Bubbles}

\begin{figure} [ht]
\centerline{\includegraphics[width=0.45\textwidth]{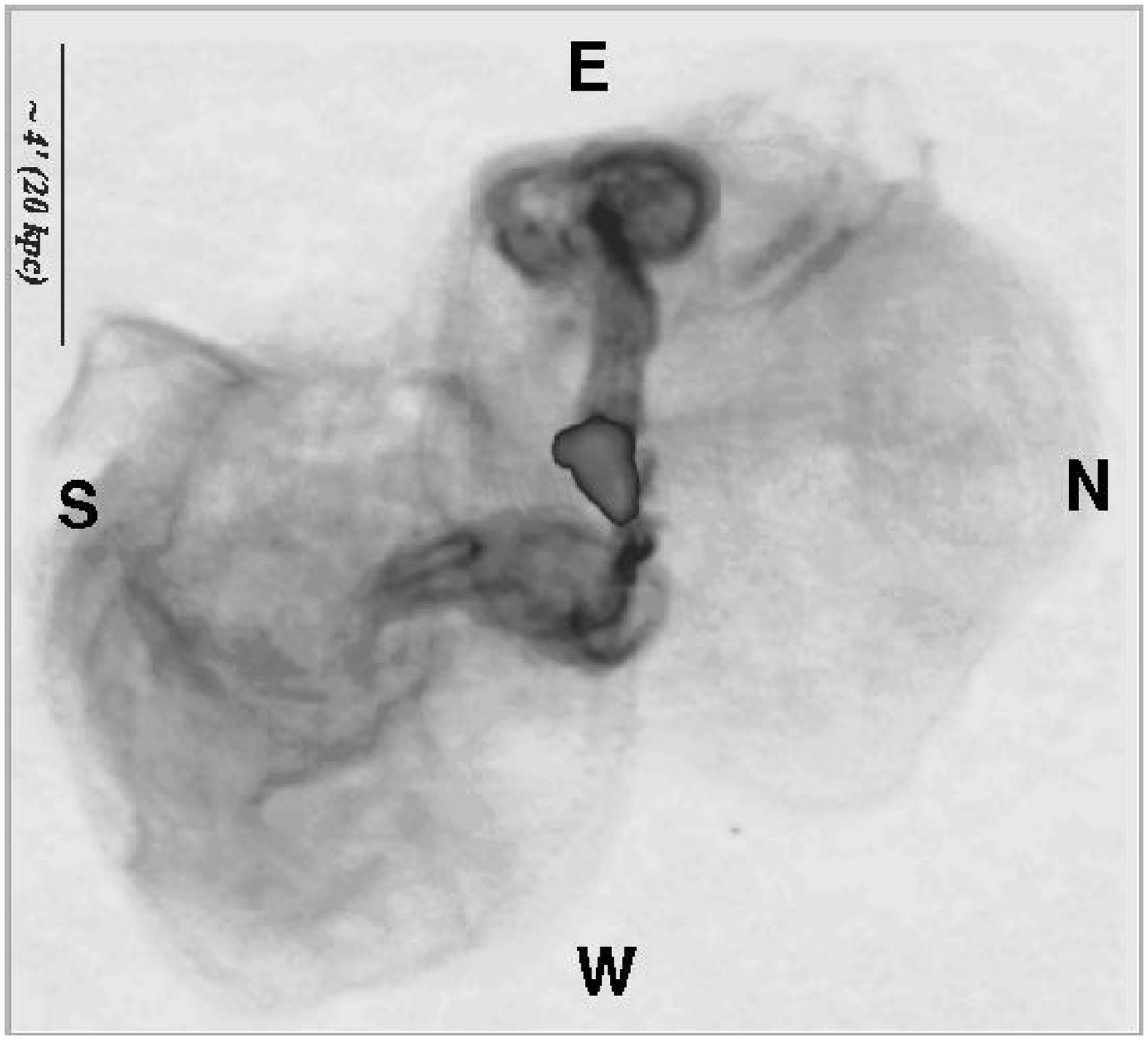}\includegraphics[width=0.45\textwidth]{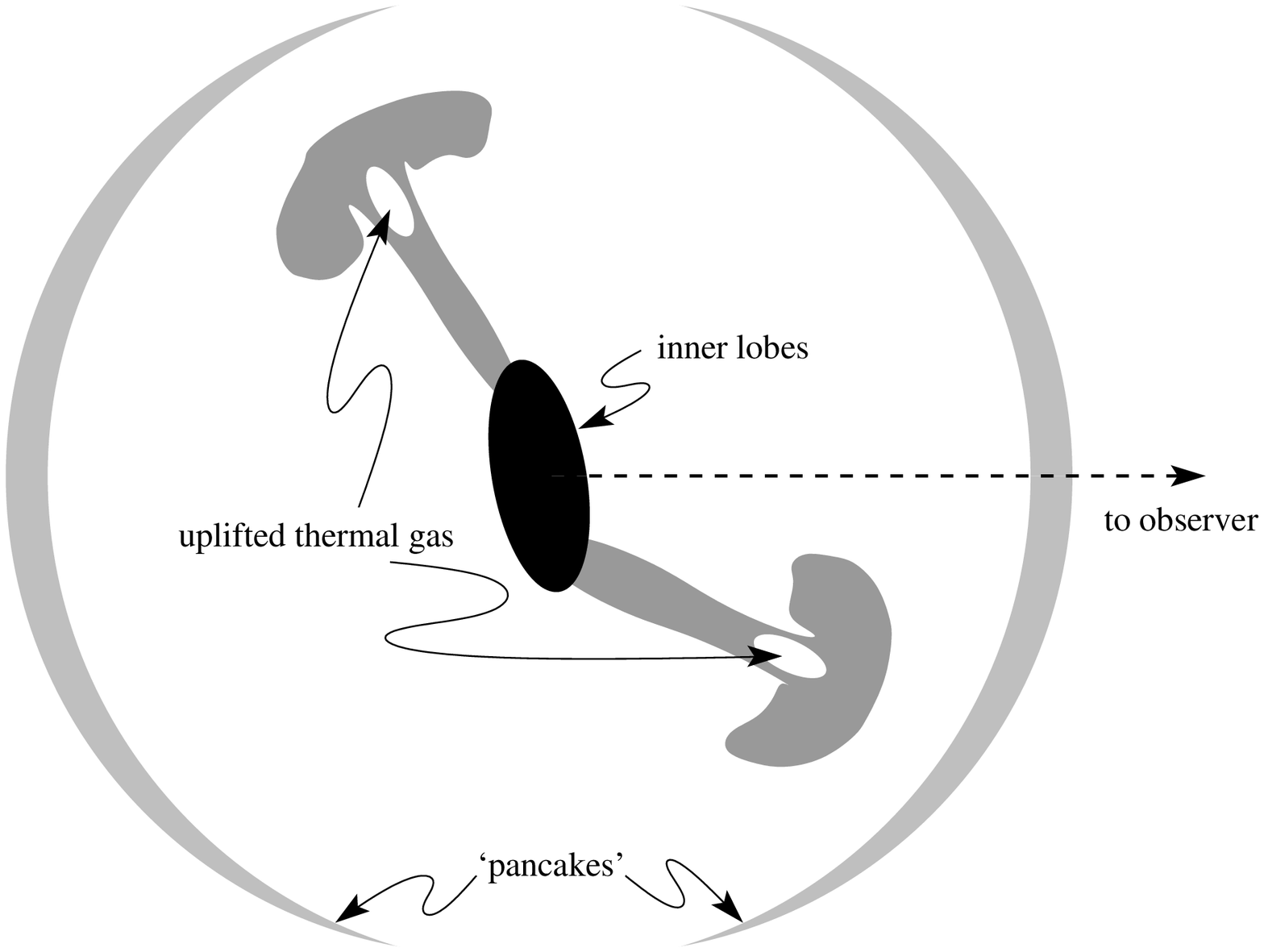}}
\caption{{\bf Left:} $14'.6 \times 16'.0$ radio map  of
M87 (North to the right, East
is up) (from Owen et al.~\cite{wrfowen2000}).
{\bf Right:} Suggested source geometry.
The central black region denotes the inner radio
lobes, the gray ``mushrooms'' correspond to buoyant bubbles already
transformed into tori, and the gray lens-shaped structures are
``pancakes'' (seen edge-on) possibly formed by older bubbles~\cite{wrfchur2001}.
}
\label{wrfm87}
\end{figure}

The 327 MHz high resolution, high dynamic range radio map (see
Fig.~\ref{wrfm87}) of M87 shows a well-defined torus-like eastern bubble
and a less well-defined western bubble, both of which are connected to
the central emission by a column, and two very faint almost circular
emission regions northeast and southwest of the
center~\cite{wrfowen2000}. The correlation between X--ray and radio
emitting features has been remarked by several
authors~\cite{wrffeigelson1987,wrfboh1995,wrfharris1999}.

\begin{figure} [!bh]

\centerline{\includegraphics[width=0.95\textwidth,bb=38  321 555 521,clip]{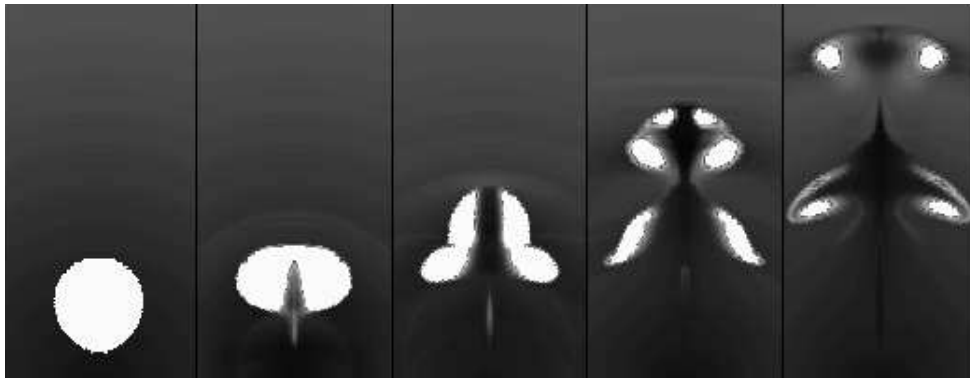}}

\caption{The gas temperature distribution at  0, 8.4, 21, 42, and 67
 Myr. Each box is $40\times20$ kpc. The temperature is coded in the 0.7
(black) to 5 keV (white) range. The hot
``radio-emitting plasma,'' which initially has a temperature on the
order of 100 keV, is white. The coldest
gas is not at the center of the cooling flow but is associated with
the rising bubble and is produced from uplifted, adiabatically
expanded gas.}
\label{wrfbubble}
\end{figure}

Motivated by the similarity in appearance between M87 and hot bubbles
rising in a gaseous atmosphere, Churazov et al. developed a simple
model of the M87 bubbles which is generally applicable to the many
bubble-like systems seen in the Chandra observations~\cite{wrfchur2001}.
Fig.~\ref{wrfbubble} shows  an initially buoyant bubble 
transform into a torus as it rises through a
galaxy or cluster atmosphere.

By entraining cool gas as it rises, the bubble produces a
characteristic ``mushroom'' appearance, similar to an atmospheric
nuclear explosion.  This qualitatively explains the correlation of the
radio and X--ray emitting plasmas and naturally accounts for the
thermal nature of the X-ray emission associated with the rising
torus~\cite{wrfboh1995,wrfboh2001,wrfbelsole2001}.  Finally, in the
last evolutionary phase, the bubble reaches a height at which the
ambient gas density equals that of the bubble.  The bubble then
expands to form a thin layer (a ``pancake'').  The large low surface
brightness features in the M87 radio map could be just such pancakes.
In the simulations
performed by Churazov et al. the buoyant bubbles behaved as expected
and did produce the features observed in both X-rays and radio for
M87. Although the exact form of the rising bubbles was sensitive to
initial conditions, the toroidal structures were a common
feature. Ambient gas was uplifted in the cluster atmosphere reducing
the effects of cooling flows and producing the
``stem'' of the mushroom that is brighter than the surrounding
regions~\cite{wrfchur2001}. Note that the XMM-Newton observation shows
that the bright X-ray columns are cool, exactly as expected in the
buoyant bubble scenario~\cite{wrfbelsole2001}.  A sketch of the
overall source structure of M87, based on the evolution of buoyant
bubbles, is shown in Fig.~\ref{wrfm87}~\cite{wrfchur2001}.

\begin{figure}
\centerline{\includegraphics[width=0.3\textwidth,bb=75 220 450
  550,clip]{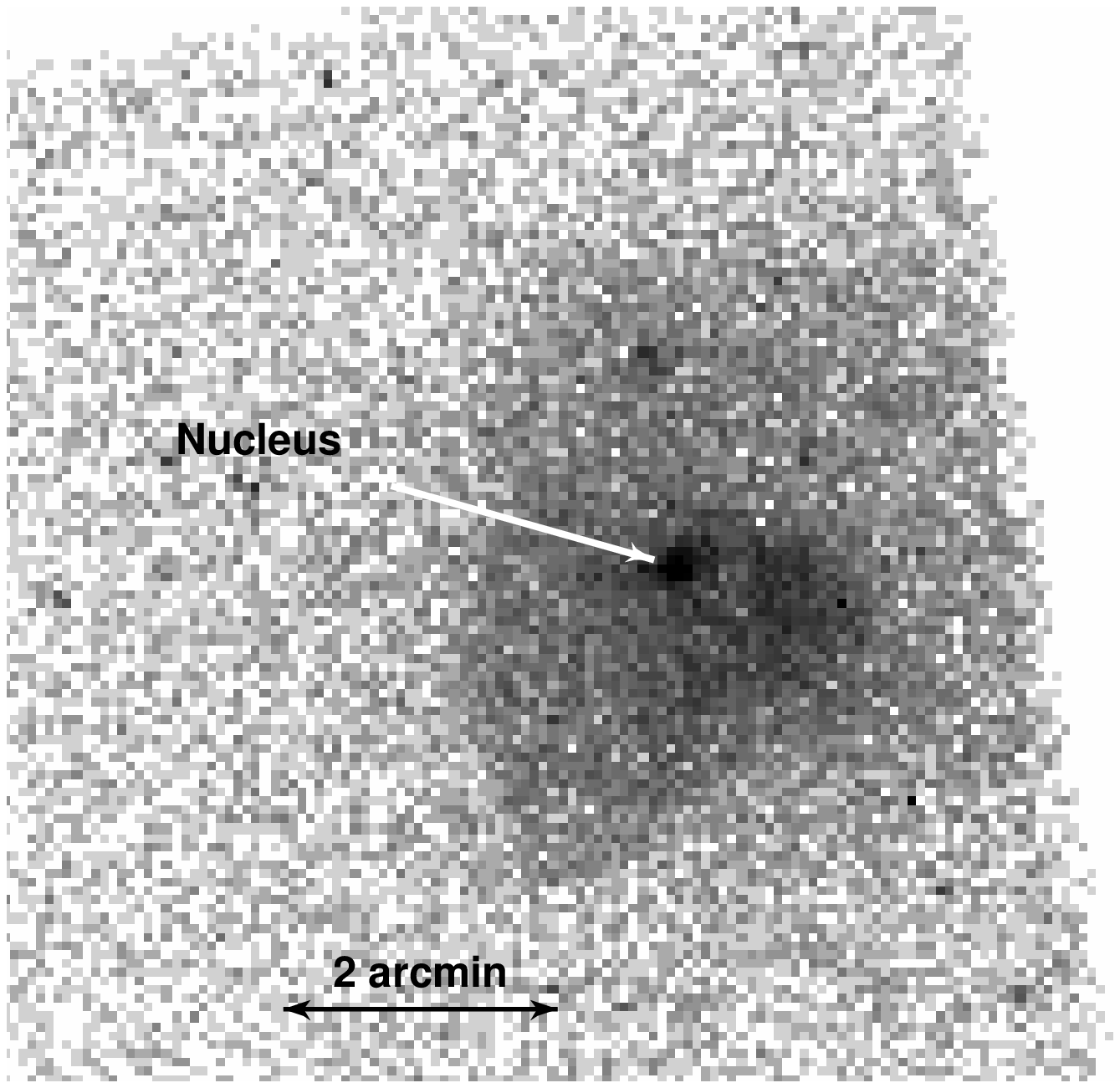}
\includegraphics[width=0.3\textwidth,bb=79 175 440 500,clip]{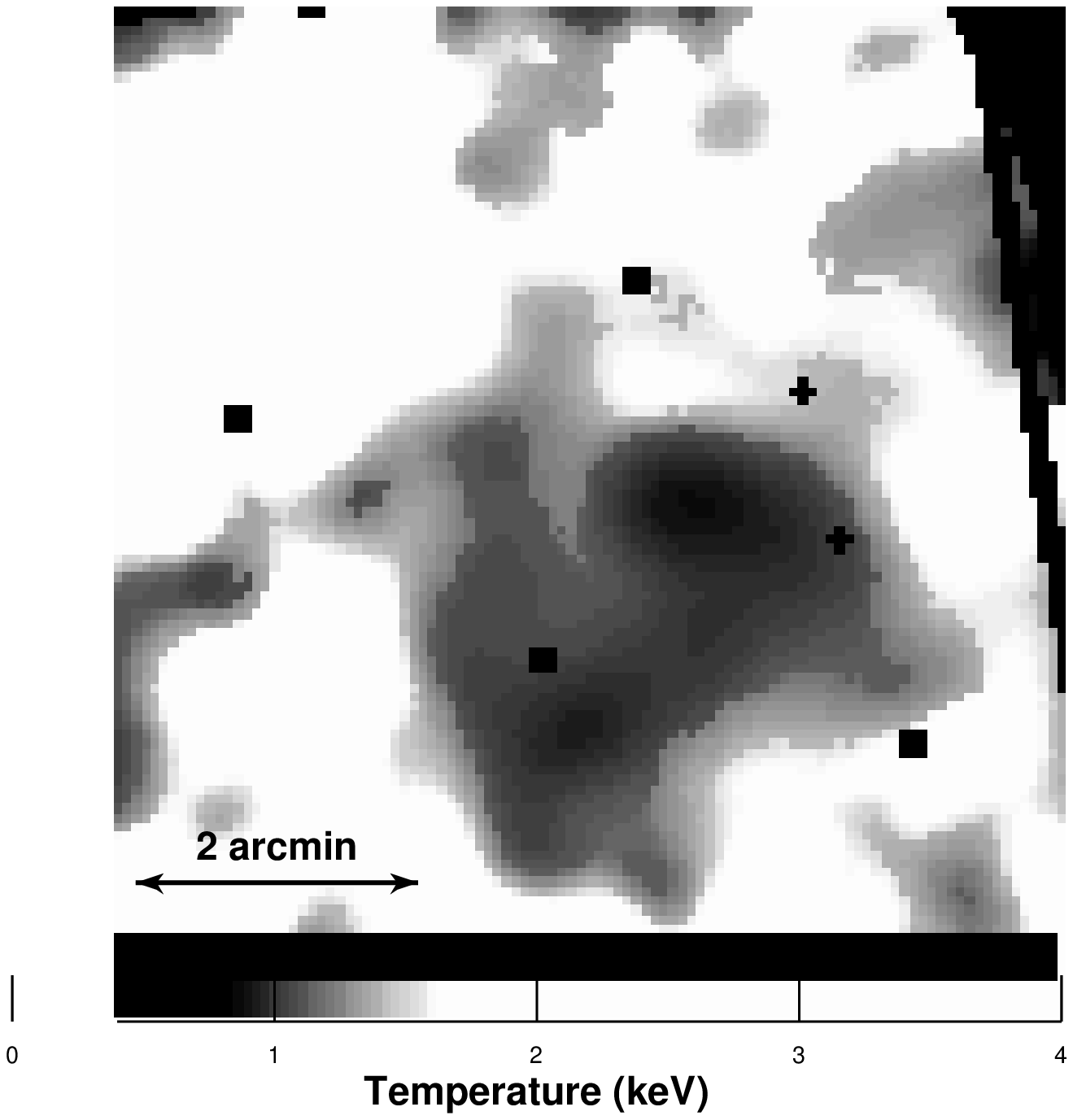}
\includegraphics[width=0.3\textwidth,bb=23 66 573 528,clip]{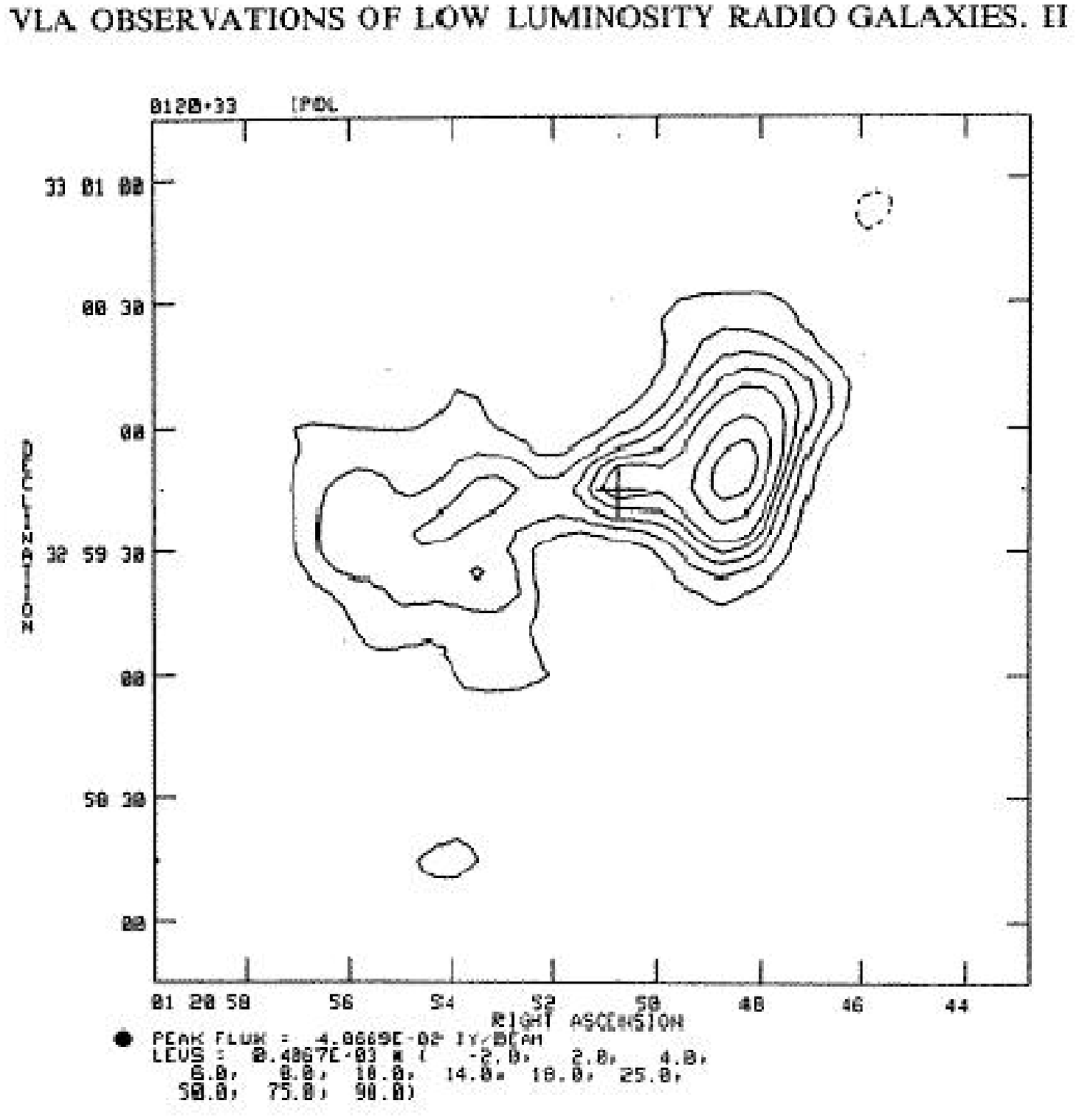}}
\caption{(\textbf{a}) The 0.5-2.0 keV surface brightness distribution of
  NGC507. (\textbf{b})  The temperature map of the central
  region of NGC507.  The galaxy center is cool as are the regions to
  the west, the north-south edge to the east, and the edge to the
  south running from northwest to southeast. (\textbf{c}) The VLA
  radio map showing the central point source, a jet emanating to the
  west and two radio lobes~\cite{wrfruiter}.  The
  depression in the X-ray surface brightness to the west of the galaxy
  peak coincides with the western radio lobe. 
  }
\label{wrfn507}
\end{figure}

\paragraph{NGC507 - the central galaxy in a group}

NGC507 is the central galaxy in a nearby ($z=0.016$) group
that has been studied extensively in
X-rays~\cite{wrfkim1995,wrfmatsumoto1997,wrfbuote1998,wrffukazawa1998}. The galaxy
is the site of a weak B2 radio source (luminosity $\sim10^{37}$ ergs
s$^{-1}$)~\cite{wrfruiter}. The Chandra X-ray image, shown in
Fig.~\ref{wrfn507}a, covers only the central, high surface
brightness emission of the group around NGC507.  The 0.5-2.0 keV surface
brightness distribution shows sharp edges to the southwest, southeast
and north, reminiscent of those in the clusters A2142 and A3667. In
addition to the edges, there are two X-ray peaks. The first, to the
east, coincides with the nucleus of NGC507. 
A second peak, $1'$ to the west has no optical
counterpart. However, comparing the X-ray and the radio map 
(Fig.~\ref{wrfn507}a, c) shows that the western radio lobe lies
precisely in the surface brightness trough between the nucleus and the
peak to the west. Thus, it seems likely that the radio lobe, probably
a buoyant bubble, has displaced X-ray emitting gas generating a trough
in the X-ray surface brightness distribution.

The origin of the peculiar sharp surface brightness discontinuities
around NGC507 is unclear. The bright emission is well fit by a thermal
model with gas temperatures near 1 keV, consistent with the mean ASCA
temperature of $1.10\pm0.05$ keV~\cite{wrfmatsumoto1997}. Detailed
spectroscopic analyses were complicated by unexpected high background
at high energies. A second Chandra observation is planned to allow
detailed spectroscopy. The emission from the central region is
resolved and hence the contribution from a central AGN is relatively
small (see Forman et al.~\cite{wrfmoriond} for additional discussion
of NGC507). Perhaps the X-ray surface brightness features arise either
from motion of NGC507 and its dark halo within the larger group
potential as suggested for the multiple edges in
clusters~\cite{wrfa1795}.

\section{Conclusions}

We did not expect the rich variety of new structures seen in the
Chandra high angular resolution
observations of clusters and early type galaxies.  Instead of
confirming our prejudices, Chandra has brought us a wealth of new
information on the interaction of radio sources with the hot gas in
both galaxy and cluster atmospheres. We see ``edges'' in many systems
with hot and cold gas in close proximity and have been able to extract
important new parameters of the ICM.  We have only
barely begun to digest the import of the Chandra cluster and galaxy
observations.  We can only expect the unexpected as Chandra
observations continue and as our understanding of how best to use this
new observatory matures.

We acknowledge support from NASA contract NAS8 39073, NASA grants
NAG5-3065 and NAG5-6749 and the Smithsonian Institution.

\centerline{\Large \bf Note Added on ZW3146}

\begin{figure} [bh]
\vspace*{0.2in}
\centerline{\includegraphics[width=0.8\textwidth]{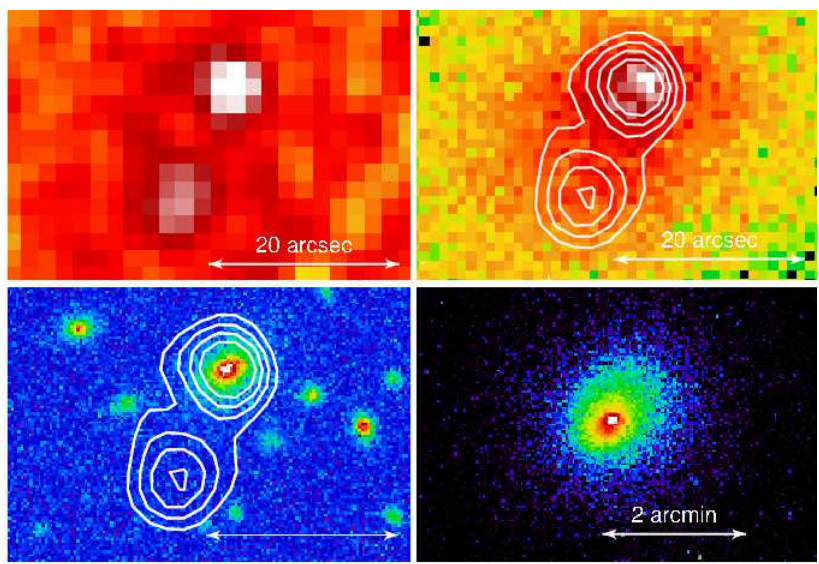}
}
\caption{(\textbf{a - upper left}) The FIRST radio image of ZW3146 shows two
radio sources with integrated fluxes of $\sim2$ mJy. (\textbf{b -
lower left}) Contours of the FIRST radio image superposed on an
optical image (S. Allen and A. Edge, private communication) shows
that the northern radio source is nearly perfectly aligned on the
center of the optical cD galaxy at the cluster center. The second
radio source may be a background object, unrelated to the cluster.
(\textbf{c - upper right}) Contours from the FIRST radio image
superposed on the Chandra full resolution ($0.492''$ pixel) image
show that the radio emission is centered within an irregular shell of X-ray
emission. (\textbf{d - lower right}) Large scale
X-ray image of ZW3146.}
\label{wrfzw3146_new}
\end{figure}

One of the puzzles relating to the X-ray observation of ZW3146 was the
apparent misalignment by a few arcseconds between the X-ray brightness
peak and the optical center of the central, dominant cD galaxy. The
solution to this apparent misalignment was resolved by the FIRST radio
image which shows two radio sources with flux densities of $\sim2$ mJy
separated by $14''$ (see Fig.~\ref{wrfzw3146_new}a). As
Fig.~\ref{wrfzw3146_new}b shows, the northern component of the pair of
radio sources is precisely aligned with the optical cD galaxy at the
cluster center.  Fig.~\ref{wrfzw3146_new}c shows that the radio
source, and therefore the center of the cD galaxy, lies at the
approximate center of irregular X-ray features.  The bright X-ray
structure northwest of the cD is elongated perpendicular to the direction
to the galaxy center. 
A second X-ray bright feature lies southeast of
the radio peak.  We suggest that the radio emitting plasma formed
a bubble that produced a trough in the X-ray surface brightness at
the galaxy center. The bubble is surrounded by an irregular shell of X-ray
emission. Thus, the brightest X-ray structures do not lie at the
galaxy center, but instead form a partial shell surrounding the radio
emitting plasma~\cite{wrfzw3146}.  Such structures are similar to the 
X-ray surface brightness enhancements seen around radio plasma bubbles in
nearby galaxies e.g., M84~\cite{wrffinoguenov2001}.

\end{document}